\documentclass[aps,prc,twocolumn,superscriptaddress,showpacs]{revtex4}

\usepackage[dvips]{graphicx} 
\usepackage{color}           
\usepackage{amsmath,amssymb,amsfonts}
\usepackage{bm}
\newcommand{\nc}{\newcommand}           
\nc{\vc}[1]     {\mbox{\boldmath $#1$}} 
\nc{\mapleft}[1]{                       
 \smash{\mathop{                      %
  \hbox to 0.90cm{\rightarrowfill} }\limits_{#1}}}

\nc{\red}[1]    {\textcolor{red}{#1}}  

\nc{\wtil}      {\widetilde}            
\nc{\bra}       {\langle}               
\nc{\ket}       {\rangle}               
\nc{\bras}[1]   {\langle#1|}            
\nc{\kets}[1]   {|#1\rangle}            
\nc{\hO}        {\hat{O}}           

\nc{\mydraft}	{\setlength{\topmargin}{-1.5cm}}
\mydraft

\begin{document}

\title{Modeling the electric dipole strength of neutron-rich $^8$He: Prediction of multineutron collective excitations}

\author{Takayuki Myo\footnote{takayuki.myo@oit.ac.jp}}
\affiliation{General Education, Faculty of Engineering, Osaka Institute of Technology, Osaka, Osaka 535-8585, Japan}
\affiliation{Research Center for Nuclear Physics (RCNP), Osaka University, Ibaraki 567-0047, Japan}

\author{Kiyoshi Kat\=o\footnote{kato@nucl.sci.hokudai.ac.jp}}
\affiliation{Nuclear Reaction Data Centre, Faculty of Science, Hokkaido University, Sapporo 060-0810, Japan}

\date{\today}

\begin{abstract}
We investigate the multineutron excitations of neutron-rich $^8$He in the electric dipole strength using a $^4$He+$n$+$n$+$n$+$n$ five-body model.
Many-body unbound states of $^8$He are obtained in the complex scaling method.
We found that the different excitation modes coexist in the dipole strength below the excitation energy of 20 MeV.
The strength below 10 MeV comes from the $^7$He+$n$ channel,
indicating the sequential breakup of $^8$He via the $^7$He resonance to $^6$He+$n$+$n$.
Above 10 MeV, the strength is obtained from many-body continuum states with strong configuration mixings,
suggesting a new collective motion of four neutrons with the dipole oscillation against $^4$He.
\end{abstract}



\maketitle 

\noindent{\mbox{\boldmath$Introduction$}}:
Physics of unstable nuclei has been developed with experiments using radioactive beams.
The neutron halo is an exotic structure of unstable nuclei near the drip-line, such as $^6$He, $^{11}$Li, and $^{11}$Be \cite{tanihata85,tanihata13}.
A common feature of unstable nuclei is a weak binding of a few excess nucleons, and many states can be observed above the particle thresholds.
This means that the spectroscopy of resonances and the responses of continuum states provide important information to understand the properties of unstable nuclei. 
The Coulomb breakup experiment is a useful tool to investigate the electric dipole response of unstable nuclei,
and may cause exotic excitation modes owing to the excess neutrons \cite{nakamura94,aumann99,nakamura06}.
The soft dipole mode, which can also be observed as a resonance, is the one of the characteristic excitations of unstable nuclei induced by the excess neutrons \cite{hansen87,ikeda92,adrich05,paar07}.
In relation to neutron-rich systems, recently tetraneutron has become an object of interest to clarify the possible existence of multineutron systems \cite{kisamori2016,duer2022}.

In neutron-rich He isotopes, $^8$He has four excess neutrons above $^4$He, which make a neutron-skin structure with a small separation energy of 3.1 MeV.
Many experiments of $^8$He have been reported \cite{korsheninnikov93,iwata00,meister02,chulkov05,mueller07,golovkov09,holl21}.
It is interesting to explore the exotic excitations of the excess neutrons and the possibility of the soft dipole mode in He isotopes experimentally \cite{lehr22} and theoretically \cite{bonaiti22,piekarewicz22}.

The lowest threshold of particle emissions in $^8$He is the $^6$He+$n$+$n$ three-body channel with the excitation energy of 2.1 MeV,
and the next open--channel is $^7$He+$n$ with the energy of 2.6 MeV.
There are experimental reports on the excited states located above the $^4$He+$n$+$n$+$n$+$n$ five-body threshold energy \cite{golovkov09,holl21}.
These facts indicate that the low-energy excitations of $^8$He bring many-body breakups to the range of the channels $^7$He+$n$, $^6$He+$n$+$n$, $^5$He+$n$+$n$+$n$, and $^4$He+$n$+$n$+$n$+$n$.

In this paper, we investigate the electric dipole strength of $^8$He in the low-energy region and
examine the new excitation modes such as the soft dipole mode of the excess neutrons oscillating against the $^4$He core. 
We employ a $^4$He+$n$+$n$+$n$+$n$ five-body cluster model and
describe many-body unbound states by using the complex scaling method (CSM) \cite{ho83,moiseyev98,aoyama06,moiseyev11,myo14a,myo20},
imposing the correct boundary conditions for decaying states.
So far, we have systematically obtained the resonances of the neutron-rich He isotopes and their mirror proton-rich unbound nuclei in the present cluster model \cite{myo10,myo12,myo21}.
We solve the motion of multineutron around the $^4$He core in the cluster orbital shell model (COSM) \cite{suzuki88,masui06,masui12}.
In the COSM, one can reproduce the threshold energies of the particle emissions in $^8$He.
This condition is important in the description of the open channels of multineutron emissions.
In the CSM, we solve the eigenvalue problem of the complex-scaled Hamiltonian and obtain the resonances explicitly.
In addition, the strength functions are described by using the Green's function with complex scaling \cite{myo98,myo01,myo07b,suzuki05,odsuren15}. 
For Coulomb breakup reactions of halo nuclei $^6$He and $^{11}$Li, we have successfully investigated the three-body breakup cross sections \cite{myo01,myo07b,kikuchi10,kikuchi13}.
In particular, we can decompose the strength function into the various channels of the direct three-body breakup and the sequential breakup via the subsystems.
Similarly for $^8$He, we clarify the effects of the various open-channels in the dipole strength with complex scaling.

\vspace*{0.2cm}
\noindent{\mbox{\boldmath$Method$}}: The five-body model of $^8$He is explained in Refs. \cite{myo10,myo12,myo21}.
\begin{figure}[b]
\centering
\includegraphics[width=8.6cm,clip]{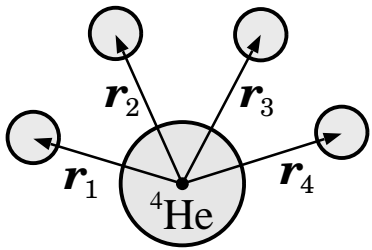}
\caption{Coordinate system of $^4$He+$n$+$n$+$n$+$n$ in the COSM.}
\label{fig:COSM}
\end{figure}
The relative coordinates of four valence neutrons measured from $^4$He are $\{\vc{r}_i\}$ with $i=1,\ldots,4$ as shown in Fig.~\ref{fig:COSM} in the COSM.
The total wave function $\Psi^J$ with spin $J$ for $^8$He is represented by the superposition of the configuration $\Psi^J_c$ as
\begin{eqnarray}
    \Psi^J
&=& \sum_c C^J_c \Psi^J_c,
    \qquad
    \Psi^J_c
~=~ \prod_{i=1}^{4} a^\dagger_{p_i}|0\rangle, 
    \label{WF0}
\end{eqnarray}
where the vacuum $|0\rangle$ is given by the $^4$He ground state with $s$-wave.
The creation operator $a^\dagger_p$ is for the single-particle state $p$ of a valence neutron above the $^4$He core in a $jj$-coupling scheme.
The index $c$ represents the set of $\{p_i\}$.
We take a summation of the available configurations in Eq.~(\ref{WF0}).
We expand the radial part of the single-particle state $p$ including the continuum states, using the Gaussian functions. 
For $1^-$ states of $^8$He, we consider the configurations with states up to the orbital angular momentum $\ell=3$ for the dipole transition.

The five-body Hamiltonian of $^8$He is the same as used in the previous studies~\cite{myo21}:
\begin{eqnarray}
	H
&=&	t_\alpha+ \sum_{i=1}^4 t_i - T_G + \sum_{i=1}^4 v^{\alpha n}_i + \sum_{i<j}^4 v^{nn}_{ij}
    \\
&=&	\sum_{i=1}^4 \left( \frac{\vc{p}^2_i}{2\mu} + v^{\alpha n}_i \right) + \sum_{i<j}^4 \left( \frac{\vc{p}_i\cdot \vc{p}_j}{4 m} + v^{nn}_{ij} \right) .
    \label{eq:Ham}
\end{eqnarray}
The kinetic energy operators $t_\alpha$, $t_i$, and $T_G$ are those of $^4$He, one valence neutron and the center-of-mass motion, respectively.
The operator $\vc{p}_i$ is the relative momentum between $^4$He and a valence neutron.
The $^4$He--neutron interaction $v^{\alpha n}$ is given by the microscopic Kanada-Kaneko-Nagata-Nomoto potential \cite{aoyama06,kanada79},
which reproduces the $^4$He-nucleon scattering data.
For neutron-neutron interaction $v^{nn}$, we use the Minnesota central potential \cite{tang78}.
In this study, we employ the 173.7 MeV of the repulsive strength of the Minnesota potential $v^{nn}$ instead of the original 200 MeV
to reproduce 0.98 MeV of the two-neutron separation energy of $^6$He \cite{myo21}.


The operator of the electric dipole ($E1$) transition in the present five-body model is expressed using the relative coordinate $\vc{r}_i$ in Fig. \ref{fig:COSM}:
\begin{eqnarray}
  \hO_{E1,\mu}&=& -\frac{eZ_\alpha}{A} \sum_{i=1}^{4} r_i Y_{1\mu}(\widehat{\vc{r}}_i), 
\end{eqnarray}
where $Z_\alpha$ is a proton number of $^4$He and $A$ is a mass number of total system.
In this study, the excitation of $^4$He is not treated and this component can contribute to the dipole strength of $^8$He
above the excitation energy of around 20 MeV.

We describe many-body unbound states using the CSM \cite{ho83,moiseyev98,moiseyev11,aoyama06,myo14a,myo20}.
In the CSM, the relative coordinates $\{\vc{r}_i\}$ are transformed using a scaling angle $\theta$ as
$\vc{r}_i \to \vc{r}_i\, e^{i\theta}$.
The Schr\"odinger equation is expressed using the complex-scaled Hamiltonian $H_\theta$ as 
\begin{eqnarray}
	H_\theta   \Psi^J_\theta
&=&     E^J_\theta \Psi^J_\theta ,
	\label{eq:eigen2}
        \\
    \Psi^J_\theta
&=& \sum_c C^J_{c,\theta} \Psi^J_c .
    \label{eq:WF_CSM}
\end{eqnarray}
We solve the eigenvalue problem of Eq.~(\ref{eq:eigen2}) and
obtain the complex-scaled wave function $\Psi^J_\theta$ in Eq.~(\ref{eq:WF_CSM}).
All of the energy eigenvalues $E^J_\theta$ are obtained and discretized for bound, resonant and continuum states.

In the wave function $\Psi^J_\theta$, the coefficients $C_{c,\theta}^J$ are the $\theta$-dependent complex numbers.
We obtain $E^J_\theta$ on a complex energy plane according to the so-called ABC theorem \cite{ABC}.
In the CSM, every Riemann branch cut is rotated down by $2\theta$ in the complex energy plane.
The branch cuts start from the threshold energies of the particle emissions and the continuum states are obtained on the corresponding $2\theta$ lines.
The energies of bound and resonant states are independent of $\theta$.
One does not use the Hermitian product according to the bi--orthogonal property of the adjoint states \cite{ho83,moiseyev98,berggren68}.

We consider the extended completeness relation (ECR) of $^8$He consisting of bound and unbound states
using the complex-scaled eigenstates $\Psi^J_\theta$ in Eq.~(\ref{eq:WF_CSM}) \cite{myo01,myo10,berggren68}.
Five-body components of $^8$He are classified into several categories as
\begin{eqnarray}
	{\bf 1}
&=&	\sum_{~\nu} \kets{\Psi^J_{\theta,\nu}}\bras{\wtil{\Psi}^J_{\theta,\nu}}
	\nonumber
	\\
&=&	\{\mbox{bound state of $^8${He}}\}
~+~	\{\mbox{resonances of $^8${He} }\}
	\nonumber
	\\
&+&	\{\mbox{two-body continua of $^7${He}$^{(*)}$+$n$}\}
	\nonumber
	\\
&+&	\{\mbox{three-body continua of $^6${He}$^{(*)}$+$n$+$n$}\}
	\nonumber
	\\
&+&	\{\mbox{four-body continua of $^5${He}$^{(*)}$+$n$+$n$+$n$}\}
	\nonumber
	\\
&+&	\{\mbox{five-body continua of $^4${He}+$n$+$n$+$n$+$n$}\} ,
\label{eq:ECR}
\end{eqnarray}
where the index $\nu$ represents the eigenstate and $\{ \Psi^J_{\theta,\nu},\wtil{\Psi}^J_{\theta,\nu} \}$ form a set of bi--orthogonal bases.

We explain the method to calculate the strength function where we omit the spin notation for simplicity. 
One defines the complex-scaled Green's function ${\cal G}^\theta(E)$,
\begin{eqnarray}
	{\cal G}^\theta(E)
&=&	\frac{ {\bf 1} }{ E-H_\theta }
~=~	\sum_{~\nu}
	\frac{|\Psi^\theta_\nu\rangle \langle \wtil{\Psi}^\theta_\nu|}{E-E_\nu^\theta} , 
	\label{eq:green1}
\end{eqnarray}
where the complex-scaled energy eigenvalue is $E_\nu^\theta$.
The dipole strength function $S_{E1}(E)$ is expressed as
\begin{eqnarray}
	S_{E1}(E)
&=&     \sum_{~\nu} S_{{E1},\nu}(E) ,
	\label{eq:strength2}
        \\
        S_{{E1},\nu}(E)
&=&    \hspace*{-0.1cm}
        -\frac1{\pi}\ {\rm Im}\left[  \frac{
	\bras{\wtil{\Psi}_0^\theta}  (\hO_{E1}^\dagger)^\theta \kets{\Psi_\nu^\theta}
	\bras{\wtil{\Psi}_\nu^\theta} \hO_{E1}^\theta          \kets{\Psi_0^\theta}
        }{E-E_\nu^\theta}
        \right],
        \nonumber\\
&&	\label{eq:strength3}
\end{eqnarray}
where $\Psi_0$ is the initial state.
One can investigate the contributions of the state $\nu$, $S_{{E1},\nu}(E)$, in the total strength $S_{E1}(E)$
using the ECR in Eq. (\ref{eq:ECR}).
The total strength function $S_{{E1}}(E)$ is positive definite.
On the other hand, the component $S_{{E1},\nu}(E)$ is not required to be positive definite,
because $S_{{E1},\nu}(E)$ is not observable.

\begin{figure}[t]
\centering
\includegraphics[width=8.6cm,clip]{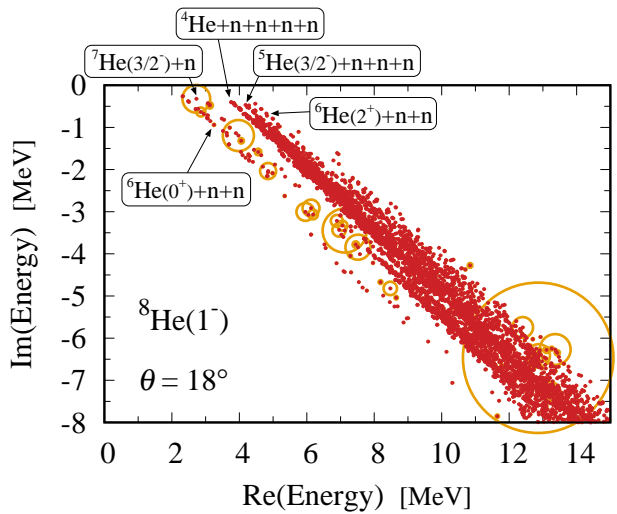}
\caption{
    Complex energy eigenvalues of the $1^-$ eigenstates of $^8$He with $\theta$=$18^\circ$ in the complex energy plane
    measured from the ground-state energy in units of MeV. Red dots indicate the complex eigenenergies.
    Some of the channels of multineutron emissions are shown.
    Size of the orange open circles corresponds to the absolute value of the real part of the dipole matrix elements
    for each eigenstate in arbitrary units.}
\label{fig:size}
\end{figure}

\vspace*{0.2cm}
\noindent{\mbox{\boldmath$Results$}}:
In Fig. \ref{fig:size}, we show the complex energy eigenvalues of $^8$He($1^-$) with dots,
measured from the ground state of $^8$He using the complex scaling with $\theta=18^\circ$,
where the ground state of $^8$He is calculated at $-3.22$ MeV measured from the $^4$He+$n$+$n$+$n$+$n$ threshold \cite{myo21}. 
The $1^-$ states are obtained along the $2\theta$ lines.
In the present calculation, the $2\theta$ lines are obtained for the branch cuts starting from the thresholds 
of $^6$He($0^+$)+$n$+$n$, $^7$He($3/2^-$)+$n$, $^4$He+$n$+$n$+$n$+$n$, $^5$He($3/2^-$)+$n$+$n$+$n$ and $^6$He($2^+$)+$n$+$n$.
The eigenstates for these eigenvalues are the discretized continuum states of every channel.
It is confirmed that no solutions of the dipole resonance are obtained in the low-energy region.
It is interesting to investigate the individual effects of the continuum states on the dipole strength function.

\begin{figure}[t]
\centering
\includegraphics[width=8.6cm,clip]{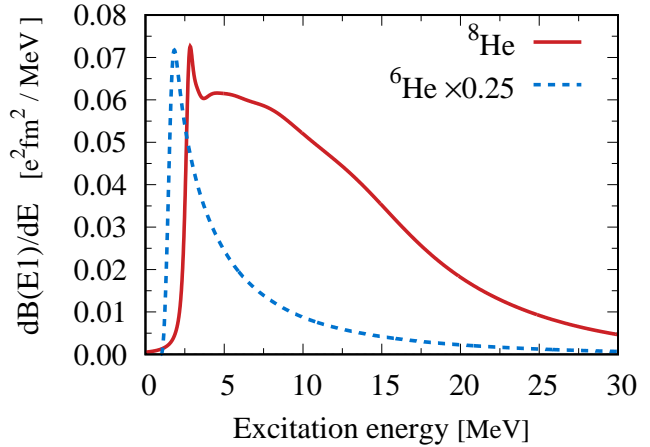}
\caption{Electric dipole strengths $dB(E1)/dE$ of $^8$He (solid) and $^6$He (dashed, multiplied by 0.25)
  as a function of the excitation energy with $\theta=18^\circ$. Units are in $e^2$fm$^2$/MeV.}
\label{fig:dist1}
\end{figure}

In Fig.~\ref{fig:dist1}, we show the electric dipole strength function $dB(E1)/dE$ of $^8$He  
corresponding to $S_{E1}(E)$ in Eq.~(\ref{eq:strength2}) as a function of the excitation energy.
It is found that the strength rapidly increases at the energy of 2 MeV and shows a spike at 2.8 MeV
and after that, a mild peak is obtained at around 5 MeV.
Above 5 MeV, the strength gradually decreases as the excitation energy increases.
It is interesting to clarify the physical origin of the energy dependence of the strength function.
For comparison, we show the electric dipole strength of $^6$He into $^4$He+$n$+$n$,
which shows the low-energy enhancement at the excitation energy of 1.9 MeV just above the open-channel of $^5$He+$n$.
This enhancement mainly comes from the $^5$He+$n$ channel with the $1s$-wave neutron \cite{myo01,kikuchi10}.

In Fig.~\ref{fig:size}, we also plot the distribution of the electric dipole matrix elements of the $1^-$ eigenstates with open circles.
The size of an open circle is proportional to the absolute value of the real part of the reduced dipole matrix element,
${\rm Re}\left\{\bras{\wtil{\Psi}^\theta_{0^+}}|\hO_{E1}^{\dagger,\theta}|\kets{\Psi^\theta_{1^-,\nu}}\bras{\wtil{\Psi}^\theta_{1^-,\nu}}|\hO^\theta_{E1}|\kets{\Psi^\theta_{0^+}}\right\}$,
which determines the magnitude of the dipole strength in the Green's function in Eq.~(\ref{eq:strength3}).
In the figure, one can easily understand the contributions of the eigenstates at the corresponding eigenenergies.
It is found that below 10 MeV,
the continuum states of the $^7$He($3/2^-$)+$n$ channel give the large matrix elements of the electric dipole strength.
On the other hand, the strengths of the other channels of $^4$He+$n$+$n$+$n$+$n$, $^5$He+$n$+$n$+$n$, and $^6$He+$n$+$n$, are minor in this energy region.
According to these results, it is meaningful to extract the component of the $^7$He+$n$ channel in the electric dipole strength function in Fig.~\ref{fig:dist1}.

In Fig.~\ref{fig:size}, we also find several $1^-$ continuum states having large electric dipole matrix elements at around 13 MeV.
In Table \ref{tab:cmp3}, we show the dominant configurations of the specific state showing the largest dipole matrix element around this energy region.
For each configuration, we take a summation of the different radial components of valence neutrons in the same orbit.
The results show a strong configuration mixing with the excitations of multineutrons from the $(p_{3/2})^4$ configuration
of the ground state of $^8$He \cite{myo21}.
This property of the continuum states can be regarded as a collective nature of four neutrons.

\begin{table}[t]
  \caption{Dominant parts of the squared amplitudes $(C_{c,\theta}^J)^2$ of the $1^-$ state of $^8$He having a signature of the collective state.
 The orbit of $\tilde{p}_{1/2}$ is orthogonal to the orbit of $p_{1/2}$.}
\label{tab:cmp3}
\centering
\begin{ruledtabular}
\begin{tabular}{c|ccc}
Configuration                             &  $(C_{c,\theta}^J)^2$ \\ \hline
 $(p_{3/2})^3(d_{5/2})$                   &  $0.313-i0.014$       \\  
 $(p_{3/2})^3(d_{3/2})$                   &  $0.311+i0.026$       \\  
 $(p_{3/2})^2(p_{1/2})(d_{5/2})$          &  $0.127-i0.017$       \\  
 $(p_{3/2})^2(p_{1/2})(d_{3/2})$          &  $0.101+i0.013$       \\  
 $(p_{1/2})^2(\tilde{p}_{1/2})(1s_{1/2})$ &  $0.098+i0.029$       \\  
 $(p_{3/2})(d_{3/2})^3$                   &  $0.055-i0.010$       \\  
 $(p_{1/2})^2(\tilde{p}_{1/2})(d_{3/2})$  &  $0.050-i0.044$       \\  
 $(p_{3/2})(p_{1/2})^2(d_{3/2})$          &  $0.030+i0.005$       \\  
 $(p_{3/2})^2(p_{1/2})(d_{3/2})$          & $-0.100+i0.065$       \\  
 $(p_{3/2})(1s_{1/2})(d_{5/2})^2$         & $-0.041+i0.001$       \\  
 $(p_{3/2})(d_{5/2})^3$                   & $-0.030-i0.013$       \\  
\end{tabular}
\end{ruledtabular}
\end{table}

\begin{figure}[th]
\centering
\includegraphics[width=8.6cm,clip]{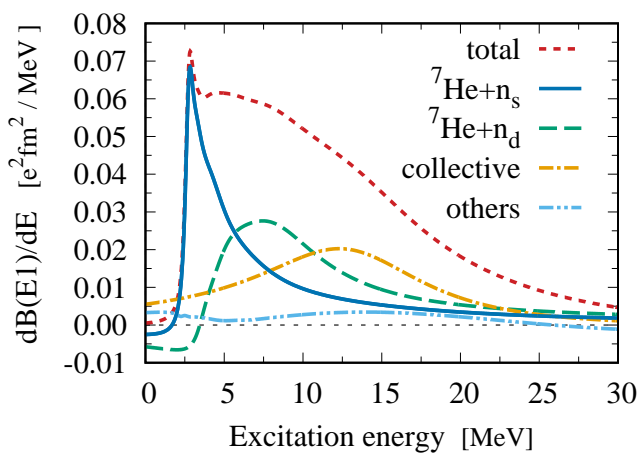}
\caption{
  Decomposition of the electric dipole strength of $^8$He as functions of the excitation energy.
  The components are
  $^7$He(3/2$^-$)+$n$ ($1s$-wave) with the blue solid line,
  $^7$He(3/2$^-$)+$n$ ($d$-wave) with the green dashed line,
  the collective states with the orange dashed dotted line,
  and the remaining other states with the light-blue dashed double-dotted line,
  in addition to the total strength with the red dotted line.}
\label{fig:cmp}
\end{figure}

In Fig.~\ref{fig:cmp}, we show the results of the decomposition of the electric dipole strength
into the components of the $^7$He+$n$ channels and the collective states using the results of Fig. \ref{fig:size}.
It is confirmed that the two-body continuum states with the $^7$He(3/2$^-$)+$n$ channels are dominant below 10 MeV.
There are two channels of the $1s$-wave (solid line) and $d$-wave (dashed line) states of the last neutron.
The $1s$-wave channel makes a spike at 2.8 MeV of the excitation energy just above the $^7$He+$n$ threshold energy,
and reflects spatial extension of the $p$-wave neutrons in the $^8$He ground state as a threshold effect.
This is a common feature of the $^5$He+$n$ case of $^6$He shown in Fig. \ref{fig:dist1} \cite{myo01}.
The $d$-wave channel makes a mild bump at around 7 MeV.
Above 10 MeV, the $^7$He+$n$ channel gradually decreases.
At around 13 MeV, the collective states (dashed dotted line) having a strong configuration mixing,
make a broad peak in the strength, which contributes to the total strength above 10 MeV.
The remaining components (dashed double-dotted line) give small contributions.

The results indicate that the sequential breakup process via $^7$He is dominant below 10 MeV
and the resonant ground state of $^7$He($3/2^-$) decays into only the $^6$He($0^+$)+$n$ state.
Above 10 MeV, another component contributes to the strength with the peak at 13 MeV
and this has a signature of the collective excitation.
This strength comes from the several non-resonant continuum states as shown in Table \ref{tab:cmp3}.
The configuration property of this component indicates the possibility of the soft dipole oscillation
of four valence neutrons against the $^4$He core. 
We investigate this component in detail by using the transition density.

We define the transition density $\rho_{E1,\nu}(r)$ for the final state $\nu$, which is the source of the electric dipole matrix element
using the reduced matrix elements as follows:
\begin{eqnarray}
  \bras{\wtil{\Psi}_{1^-,\nu}}|\hO_{E1}|\kets{\Psi_{0^+}}
  &=& \int_0^\infty \rho_{E1,\nu}(r)\, r^2 dr ,
\end{eqnarray}
where $r$ is the relative distance between $^4$He and a neutron.
In order to discuss the spatial property of $\rho_{E1,\nu}(r)$ directly,
we calculate this quantity without the complex scaling.

\begin{figure}[b]
\centering
\includegraphics[width=8.6cm,clip]{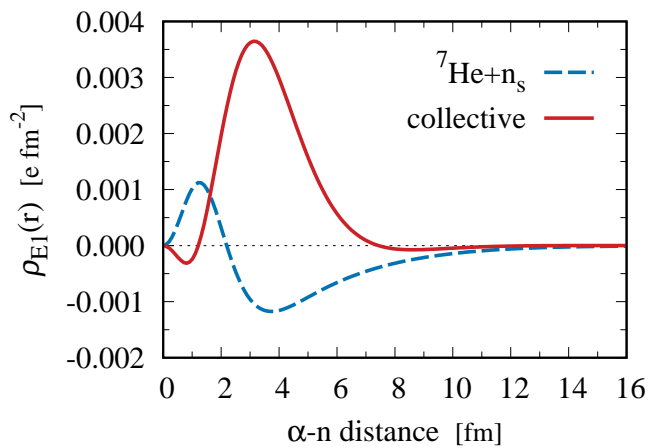}
\caption{
  Transition densities of electric dipole matrix elements of two specific $1^-$ states of $^8$He as functions of the $^4$He--neutron distance:
  the $^7$He+$n$ ($1s$-wave) channel with blue dashed line, and the collective state with red solid line.}
\label{fig:dns}
\end{figure}

We choose the state having the $^7$He+$n$ ($1s$-wave) channel with a large electric dipole matrix element,
located near the $^7$He+$n$ threshold energy.
In this state, the last neutron in $^8$He is excited from the $p$-wave to the $1s$-wave.
In Fig.~\ref{fig:dns}, we show the results of the electric dipole transition density of this component with a blue dashed line and
the long tail behavior is confirmed in the distribution.
This indicates the separation of one neutron from $^8$He by the dipole transition.
The other case is the collective state at around 13 MeV with strong configuration mixing. 
The dipole transition density with the red solid line shows a large strength at 3 fm and is distributed within the range of 2 to 6 fm.
This distribution does not show the long tail and is spatially confined. 
These properties indicates intuitively the collective motion of four valence neutrons ($4n$) in $^8$He and
can be regarded as a soft dipole mode of $4n$ against the $^4$He core \cite{ikeda92}.
We have also examined whether this collective state can be a five-body resonance, but
currently we have not obtained the specific resonance eigenenergy separated from other continuum states in the complex scaling
with a large $\theta$.
As a result, the collective excitations contribute to the electric dipole strength of $^8$He, 
and this new mode is not clearly seen in the electric dipole responses of two-neutron halo nuclei of $^6$He and $^{11}$Li
\cite{myo01,myo07b,kikuchi10,kikuchi13}.

We naively discuss the energy of 13 MeV for the collective excitation in $^8$He from the viewpoint of the relative motion between the $^4$He core and $4n$.
In the ground state, the relative distance between $^4$He and $4n$ is obtained to be 2.05 fm.
When we assume the $(0p)^4$ configuration of $4n$ in the harmonic oscillator shell model,
this condition leads to the $1s$ orbit of the relative wave function between $^4$He and $4n$ with the length parameter of 1.10 fm.
This length leads to $\hbar\omega=17$ MeV and this energy causes the dipole excitation of the relative motion and is close to 13 MeV,
but, higher by 4 MeV, which may come from a continuum effect and a deviation from the two-body picture.

\vspace*{0.2cm}
\noindent{\mbox{\boldmath$Summary$}}:
We investigated the electric dipole strength of neutron-rich $^8$He using the $^4$He+$n$+$n$+$n$+$n$ five-body cluster model.
We use the complex scaling method to obtain many-body unbound states of $^8$He.
The electric dipole strength of $^8$He shows the low-energy enhancement and has two components:
(1) Two-body continuum states consisting of the resonance of $^7$He and a neutron below 10 MeV of the excitation energy.
This is the nature of the single-particle excitations and leads to the sequential breakup of $^8$He via $^7$He to $^6$He+$n$+$n$.
(2) The collective excitations of four neutrons with strong configuration mixing, which is enhanced at 13 MeV. 
This exotic structure indicates a possibility of the soft dipole mode consisting of $^4$He and four neutrons.

The properties of the dipole strength of $^8$He show the coexistence of the different excitation modes
due to the excess neutrons, which can be general in neutron-rich nuclei
and would be confirmed in experiments such as the Coulomb breakup reaction \cite{lehr22}.
The variety of the multineutron excitations is an interesting aspect of neutron-rich nuclei.

\vspace*{0.2cm}
\noindent{\mbox{\boldmath$Acknowledgments$}}:
This work was supported by JSPS KAKENHI Grants No. JP18K03660, No. JP20K03962, and No. JP22K03643.
Numerical calculations were partly achieved through the use of SQUID at the Cybermedia Center, Osaka University.


\section*{References}
\def\JL#1#2#3#4{ {{\rm #1}} \textbf{#2}, #4 (#3)}  
\nc{\PR}[3]     {\JL{Phys. Rev.}{#1}{#2}{#3}}
\nc{\PRC}[3]    {\JL{Phys. Rev.~C}{#1}{#2}{#3}}
\nc{\PRA}[3]    {\JL{Phys. Rev.~A}{#1}{#2}{#3}}
\nc{\PRL}[3]    {\JL{Phys. Rev. Lett.}{#1}{#2}{#3}}
\nc{\NP}[3]     {\JL{Nucl. Phys.}{#1}{#2}{#3}}
\nc{\NPA}[3]    {\JL{Nucl. Phys.}{A#1}{#2}{#3}}
\nc{\PL}[3]     {\JL{Phys. Lett.}{#1}{#2}{#3}}
\nc{\PLB}[3]    {\JL{Phys. Lett.~B}{#1}{#2}{#3}}
\nc{\PTP}[3]    {\JL{Prog. Theor. Phys.}{#1}{#2}{#3}}
\nc{\PTPS}[3]   {\JL{Prog. Theor. Phys. Suppl.}{#1}{#2}{#3}}
\nc{\PRep}[3]   {\JL{Phys. Rep.}{#1}{#2}{#3}}
\nc{\JP}[3]     {\JL{J. of Phys.}{#1}{#2}{#3}}
\nc{\PPNP}[3]   {\JL{Prog. Part. Nucl. Phys.}{#1}{#2}{#3}}
\nc{\PTEP}[3]   {\JL{Prog. Theor. Exp. Phys.}{#1}{#2}{#3}}
\nc{\andvol}[3] {{\it ibid.}\JL{}{#1}{#2}{#3}}

\end{document}